\documentstyle[prl,aps,epsf]{revtex}
\newcommand{\lapx}{\mbox{\raisebox{-4pt}{$\,\buildrel<\over\sim\,$}}} 
\newcommand{\gapx}{\mbox{\raisebox{-4pt}{$\,\buildrel>\over\sim\,$}}} 
\begin{document} 
\draft 
\title{Relieving the fermionic and the dynamical sign problem:
Multilevel Blocking Monte Carlo Simulations} 
\author{R.~Egger$^1$ and C.H. Mak$^2$} 
\address{${}^1$Institut f\"ur Theoretische Physik, 
Heinrich-Heine-Universit\"at, D-40225 D\"usseldorf, Germany\\
${}^2$Department of Chemistry, University of Southern California,
Los Angeles, CA 90089-0482, USA} 
\date{Lecture notes for Kerkrade Winter school} 
\maketitle 

\begin{abstract}
This article gives an introduction
to the multilevel blocking (MLB) approach to both the fermion and
the dynamical sign problem in path-integral Monte Carlo simulations.
MLB is able to substantially relieve the sign problem in many
situations.
Besides an exposition of the method, its accuracy  and several potential
pitfalls are discussed, providing guidelines for the proper choice
of certain MLB parameters. 
Simulation results are shown for strongly interacting 
electrons in a 2D parabolic quantum dot,  the real-time dynamics
of several simple model systems, and the
dissipative two-state dynamics (spin-boson problem).
\end{abstract}

\pacs{}

\section{Introduction: The sign problem}

Quantum Monte Carlo (QMC) techniques are among the most powerful
and versatile methods for the computer simulation of
a large variety of interesting quantum systems
encountered in physics, chemistry or biology \cite{qmcgeneral}.
In particular, QMC is capable of 
delivering numerically exact results. 
Despite the great potential of this method, there
are several restrictions and handicaps inherent to
all QMC techniques, the perhaps most pressing one 
being due to the {\sl  sign problem} \cite{loh}.
There are various sign problems, namely the
{\sl fermionic sign problem}  encountered in
 equilibrium (imaginary-time)  simulations of 
strongly correlated many-fermion systems, 
and the {\sl dynamical sign problem} in real-time
(dynamical) simulations, which already
shows up for a single particle.  Unfortunately, 
apart from variational or approximate treatments (such as the
fixed-node approximation), a completely general and 
totally satisfactory solution to the sign problem in QMC 
simulations is still lacking.  Nevertheless, over
the past few years considerable and substantial progress has been achieved
along several different lines without introducing
approximations into the QMC scheme, see, for instance,
Refs.~\cite{mlb,wiese,kalos}. 
 
In these notes focus is put on one specific class
of QMC methods called {\sl Path-integral Monte Carlo (PIMC)}.
PIMC is based on a discretized path-integral representation of the quantities
of interest.  
The sign problem then arises when different paths that
contribute to averages carry different sign (or complex-valued phase).
For instance, for the fermionic sign problem, 
as a consequence of exchange, one typically has to deal with
determinants, so that
non-positive-definite fermionic density matrix elements arise. 
The sign cancellations arising from sampling fermion paths then manifest
themselves as a signal-to-noise ratio, $\eta\sim
\exp(-N\beta E_0)$, that vanishes exponentially
with both particle number $N$ and inverse temperature $\beta=1/k_B T$;
here $E_0$ is a system-dependent energy scale.
The small signal surviving the interference of many fermion paths
is then inevitably lost in the large background noise of the
stochastic simulation. 
Similarly, when studying the real-time dynamics of even a single
particle, the quantum-mechanical time evolution operator
$\exp(-i H t/\hbar)$ attaches a complex-valued phase to each
quantum path, which in turn gives rise to the dynamical sign problem.
Again the signal-to-noise ratio will vanish exponentially,
$\eta\sim \exp(-t^*/\tau_0)$, where $t^*$ is the maximum real time
under study and $\tau_0$ a system- and implementation-specific 
characteristic time scale.  The exponential scaling is typical
for the ``naive'' approach, where one simple uses the absolute
value of the complex-valued  weight function for the MC sampling,
and includes the phase information in the accumulation.

In Ref.~\cite{mlb} a general scheme for tackling the sign problem
in PIMC simulations was proposed. The method has been applied 
to interacting electrons in a quantum dot~\cite{egger99} and to the 
real-time dynamics of simple few-degrees-of-freedom
 systems \cite{rtmlb}.   The generalization of the algorithm
to the case of effective actions --- which are long-ranged along
the Trotter or real-time direction ---  along with the application
to the dissipative two-state (spin-boson) dynamics
has been given in Refs.~\cite{egger1,lothar}.
This {\sl multilevel blocking} (MLB) approach
represents the systematic
implementation of a simple {\em blocking strategy}.
The blocking strategy states that by sampling
groups of paths (``blocks'') at the same time, 
the sign problem can always be reduced compared to 
sampling single paths as would be done normally; 
for a proof, see Sec.~\ref{block} below. 
By suitably bunching paths together
into sufficiently small blocks,
 the sign cancellations among paths within the same block 
can be accounted for without the sign problem, simply because
there is no sign problem for a sufficiently small system.
The MLB approach is based on a recursive implementation
of this idea,  i.e.~after building ``elementary'' blocks, 
new blocks are formed out of these, and the process is
then iterated.  This leads to a powerful 
implementation of the blocking strategy.
This algorithm is able to turn the exponential severity
of the sign problem into an ``only'' algebraic one. 
This is still difficult enough but in practice
implies that significantly larger systems (lower temperature,
longer real time) can be studied by MLB-PIMC than under
the naive PIMC approach. 
Nevertheless, it should be stressed that MLB is  definitely {\sl not}\ a
black-box scheme.  There are several potential pitfalls related to 
incorrect or inadequate choices of certain MLB parameters, 
and one needs to be quite careful in applying this technique \cite{mischa}.
Given some experience, however, it represents
a powerful handle to relieve the sign problem, with 
potential for additional improvement.

The plan for the remainder of these notes is as follows.
In Sec.~\ref{mlbsec} the MLB approach is discussed in detail,
first in an intuitive way in Sec.~\ref{block},
and subsequently on a more formal or technical
level in Sec.~\ref{mlbform} for fermions.  The modifications for
real-time simulations are summarized in Sec.~\ref{realtime}.
This is then followed by a discussion of the performance and the
accuracy of this algorithm in Sec.~\ref{performance}.
In Sec.~\ref{effact}, the generalization to effective-action
problems is outlined.
To demonstrate the power of this approach, 
MLB results are presented in Sec.~\ref{appl}.
As an example for the fermionic sign problem, low-temperature
simulation results for strongly correlated 
electrons  in a 2D quantum dot 
are shown in Sec.~\ref{appl:dot}.
The remainder of that section is then concerned with the 
dynamical sign problem.  After presenting results 
for several simple model systems in Sec.~\ref{appl:simple}, 
the dynamics of the dissipative two-state system
is discussed in Sec.~\ref{appl:sb}.
Finally, some conclusions can be found in Sec.~\ref{conc}.

\section{Multilevel blocking (MLB) approach}
\label{mlbsec}

Before diving into the details of the MLB approach, 
the underlying idea  (``blocking strategy'') will
be explained, focusing for simplicity on 
fermionic imaginary-time simulations. 
For those interested in working with this method, 
technical details and practical guidelines are provided in Secs.~\ref{mlbform} 
to \ref{performance}.
In the last part the generalization of MLB
to PIMC simulations of the effective-action type is described.

\subsection{Blocking strategy} \label{block}

Let us consider a many-fermion system whose state is described by a set of
quantum numbers $\vec{r}$ denoting, e.g.~the positions and spins 
of {\em all}\, particles.  These quantum numbers can correspond to
electrons living on a lattice or in continuous space.
For notational simplicity, we focus
on calculating the equilibrium expectation value of a diagonal operator
or correlation function
(this can be easily generalized), 
\begin{equation} \label{fir}
\langle A\rangle = \frac{\sum_{\vec{r}} A(\vec{r})
\rho(\vec{r},\vec{r})}{\sum_{\vec{r}} 
\rho(\vec{r},\vec{r})} \;,
\end{equation}
where $\sum_{\vec{r}}$ represents either a summation for the case of a 
discrete system or an integration for a continuous system, and
$\rho(\vec{r},\vec{r}')$ denotes the (reduced) density matrix 
of the system.
In PIMC applications, imaginary time is discretized into $P$ slices of 
length $\epsilon=\beta/P$.
Inserting complete sets at each slice $m=1,\ldots,P$,  
and denoting the corresponding configuration on slice $m$ 
by $\vec{r}_m$, the diagonal elements of the density matrix at
$\vec{r}=\vec{r}_P$ entering Eq.~(\ref{fir}) are:
\begin{equation}\label{rho}
\rho(\vec{r}_P,\vec{r}_P) = \sum_{\vec{r}_1,\ldots, \vec{r}_{P-1}}
\prod_{m=1}^P \langle \vec{r}_{m+1}| e^{-\epsilon H}
|\vec{r}_m \rangle \;.
\end{equation}
Of course, periodic boundary conditions have to be
enforced here, $\vec{r}_{P+1}=\vec{r}_1$.
To proceed, one then has to construct 
accurate analytical approximations for the
short-time propagator.
This formulation of the problem excludes 
effective actions such as those arising from an integration over the fermions
using the Hubbard-Stratonovich trans\-for\-mation \cite{loh},
since that generally leads to long-ranged imaginary-time interactions.
The MLB approach suitable for such a situation \cite{egger1} is
described below in Sec.~\ref{effact}.

Since we are dealing with a many-fermion system, the short-time
propagators need to be antisymmetrized, leading to the
appearance of determinants causing the sign problem.
Strictly speaking, the antisymmetrization has to be
done only on one time slice, but the 
intrinsic sign problem is much better behaved if one antisymmetrizes
on all time slices.  Choosing the absolute value
of the product of the short-time propagators in Eq.~(\ref{rho})
as the positive definite MC weight function $P[X]$, one has to
keep the sign $\Phi[X]$ associated with
a particular discretized path $X=(\vec{r}_1,\ldots,\vec{r}_P)$
during the accumulation procedure,
\begin{equation}
\langle A \rangle =  \frac{\sum_{X} P[X] \Phi[X] A[X]} 
{\sum_{X} P[X] \Phi[X]} \;.
\end{equation}
Assuming that there are no further exclusivity problems in the
 numerator so that $A[X]$ is well-behaved, one can then analyze
the sign problem in terms of the variance of the
denominator, 
\begin{equation} \label{var}
\sigma^2 \approx \frac{1}{N_s} \left( \langle \Phi^2 \rangle
- \langle \Phi \rangle^2 \right)   \;,
\end{equation}
where $N_s$ is the number of MC samples taken and stochastic averages are 
calculated with $P$ as the weight function. 
For the fermion sign problem, where $\Phi=\pm 1$ and hence
$\langle \Phi^2\rangle = 1$,  the variance of the 
signal is controlled by the size of $|\langle\Phi\rangle|$.

Remarkably, one can achieve considerable progress by 
{\em blocking paths together}.
Instead of sampling single paths along the MC trajectory, one
can consider sampling sets of paths called blocks.
Under such a blocking operation,
the stochastic estimate for $\langle A \rangle$
takes the form
\begin{equation} \label{start2}
\langle A \rangle =  \frac{\sum_{B} \left(
\sum_{X\in B} P[X] \Phi[X] A[X] \right)} 
{\sum_{B} \left(\sum_{X\in B} P[X] \Phi[X] \right)} \;,
\end{equation}
where one first sums over the configurations belonging to a 
block $B$ in a way that is not affected by the sign problem, and then
stochastically sums over the blocks.  The summation within a block
must therefore be done non-stochastically, or alternatively the
block size must be chosen sufficiently small.
Of course, there is considerable freedom in how to choose this
blocking.

Let us analyze the variance $\sigma^{\prime 2}$
of the denominator of Eq.~(\ref{start2}). First define new 
sampling functions 
in terms of the blocks which are then sampled stochastically,
\begin{equation}
 P'[B] = \left| \sum_{X \in B} P[X] \Phi[X] \right| \;,\quad
\Phi'[B] ={\rm sgn} \left(\sum_{X \in B} P[X] \Phi[X]\right) \;.
\end{equation}
Rewriting the average sign in the new representation, i.e.~using 
$P'[B]$ as the weight, then inserting
the definition of $P'$ and $\Phi'$ in the numerator,
\[
\langle \Phi'[B] \rangle = 
\frac{\sum_{B} P'[B] \Phi'[B]}{ \sum_B P'[B]}
= \frac{\sum_{X} P[X] \Phi [X]}{ \sum_B P'[B]} \;,
\]
and comparing to the average sign in the standard 
representation using $P[X]$ as the weight, one finds
\[
\frac{|\langle \Phi' \rangle|}{ |\langle \Phi \rangle|} =
\frac{\sum_{X} P[X]} {\sum_{B} P'[B]} \;.
\]
By virtue of the Schwarz inequality,  
\[
\sum_{B} P'[B] = \sum_{B} \left| \sum_{X\in B} 
P[X] \Phi[X] \right|\leq \sum_{B} \left| \sum_{X\in 
B} P[X] \right|  = \sum_{X} P[X]\;,
\]
it follows that {\em for any kind of blocking}, the average sign
improves,
$|\langle\Phi'\rangle|
\geq | \langle \Phi \rangle|$.
Furthermore, since $\langle \Phi^{\prime 2} \rangle
= \langle \Phi^{2} \rangle = 1$, Eq.~(\ref{var}) implies that 
\begin{equation} \label{eqi} 
\sigma^{\prime\,2} \leq \sigma^2\;,
\end{equation} 
and hence {\em the signal-to-noise ratio is always improved upon 
blocking configurations together}.  Clearly,
the worst blocking one could possibly 
choose would be to group the configurations into two separate
blocks, one collecting all paths with
positive sign and the other with negative sign. In this 
case, blocking yields no improvement whatsoever, and the ``$\leq$''
becomes ``$=$'' in Eq.~(\ref{eqi}).  It is apparent from Eq.~(\ref{eqi}) 
that the blocking strategy provides a systematic handle to 
reduce the sign problem.   In our realization of the blocking
strategy, a block is defined by all paths that differ only at
one time slice, i.e.~only $\vec{r}_m$ is updated with all
other $\vec{r}_{n\neq m}$ being frozen.

A direct implementation of the blocking strategy 
does indeed improve the sign problem but will not remove its exponential
severity.  The reason is simply that for a sufficiently large
system, there will be too many blocks, and once the 
signals coming from these blocks are allowed to interfere,
one again runs into the sign problem, albeit with a smaller
scale $E_0$ entering the signal-to-noise ratio.
The resolution to this problem comes from
the multilevel blocking (MLB) approach~\cite{mlb} 
where one {\em applies the
blocking strategy in a recursive manner to the blocks} again.
In a sense, new blocks containing a sufficiently 
small number of elementary ones are formed, and this process  is repeated
until only one block is left.  Each step of this hierarchy 
is called {\em level}\ in what follows.
Blocks are then defined living on these
{\em levels}, and after taking care of the sign cancellations
within all blocks on a given (fine) level, the resulting sign 
information is transferred to the next (coarser) level. 
On each step, the blocking strategy ensures that no sign
problem occurs {\em provided one has chosen sufficiently
small block sizes}.
  By doing this recursively, the sign problem on all the coarser levels 
can be handled in the same manner.  
It is then possible to proceed without 
numerical instabilities from the bottom up to the top level.
The cancellations arising at the top level will create a
sign problem again, which is however strongly reduced.
As is argued below, the resulting sign problem is characterized by
an only algebraic severity.

In many ways, the MLB idea is related to the renormalization
group approach.  But instead of integrating out information
on fine levels, sign cancellations are ``synthesized''
within a given level and subsequently their effects are 
transferred to coarser levels. 
While the renormalization group filters out information
judged ``relevant'' and then ignores the ``irrelevant'' part,
no such approximation is introduced here.
Therefore our approach is actually closer in spirit to 
the multi-grid algorithm \cite{mg}.
The technical implementation of MLB is discussed next
following Ref.~\cite{mlb}.

\begin{figure}
\epsfysize=10cm
\epsffile{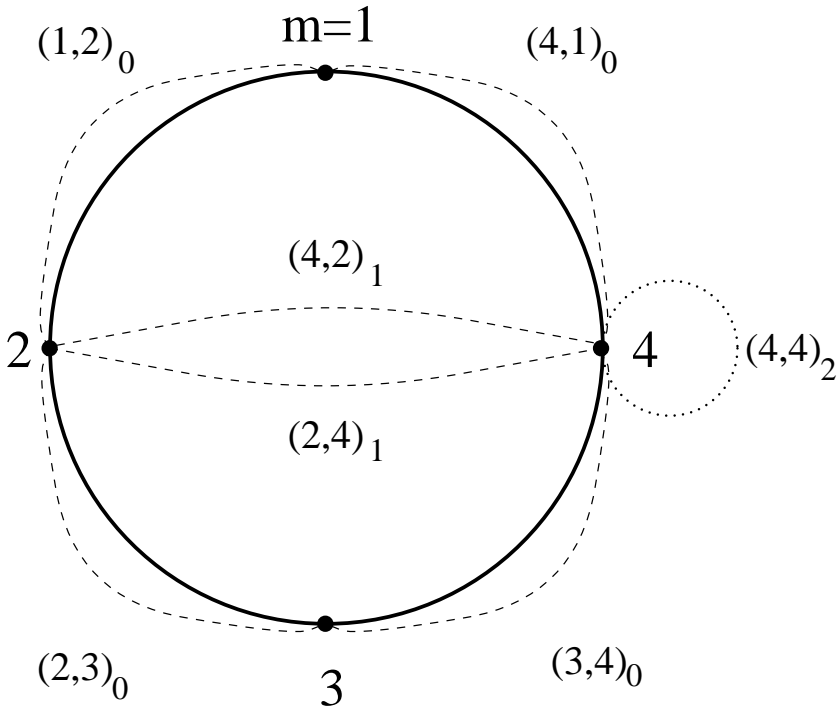}
\caption[]{\label{fig0}
Levels for $L=2$ ($P=4$). Imaginary time flows along the 
circle (solid curve), and the slices $m=1,2,3,4$ are distributed among
the three levels: The finest level $\ell=0$
contains  $m=1,3$, level $\ell=1$ contains $m=2$, and
 $\ell=2$ contains $m=4$. Level-$\ell$
bonds are indicated by dashed and dotted lines.  }
\end{figure}

\subsection{Systematic implementation: MLB}\label{mlbform}

To keep notation simple, the slice index $m$  is used
as a shorthand notation for the quantum numbers $\bbox{r}_m$.
From Eq.~(\ref{rho}) the {\em level-0 bonds},
which are simply the short-time propagators, then follow in the form:
\begin{equation}\label{short}
(m,m+1)_0 = \langle \bbox{r}_{m+1}| e^{-\epsilon H}
|\bbox{r}_m \rangle \;.
\end{equation}
Next the different levels $0\leq \ell \leq L$, where $L$
defines the Trotter number $P=2^L$, have to be specified.
Each slice $m$ belongs to a unique level $\ell$, such that $m=(2j+1)2^\ell$
and $j$ is a nonnegative integer.  For instance, the slices
$m=1,3,5,\cdots, P-1$ belong
to $\ell=0$,  $m=2,6,10,\cdots, P-2$ belong to $\ell=1$,
etc., such that there are ${\cal N}_\ell=  2^{L-\ell-1}$ 
(but ${\cal N}_L=1$)
different slices on level $\ell$, see Fig.~\ref{fig0}.
An elementary blocking is achieved by grouping together
configurations that differ only at slice $m$, so 
only $\bbox{r}_m$ varies in that block while all
$\bbox{r}_{m^\prime\neq m}$ remain fixed.
Sampling on level $\ell$ therefore 
extends over configurations $\{\bbox{r}_m\}$ living on the 
${\cal N}_\ell$ different slices.
In the MLB scheme, one moves recursively from  
the finest ($\ell=0$) up to the coarsest level ($\ell=L$), and
the measurement of the diagonal operator is done only at the
top level using the configuration $\bbox{r}_P$.

A MC sweep starts by changing only configurations
associated with the slices on level $\ell =0$
 according to the standard weight
\begin{equation}
{\cal P}_0 = |(1,2)_0 (2,3)_0 \cdots (P,1)_0| \;,
\end{equation}
generating a MC trajectory containing $K$ samples for each slice 
on level $\ell=0$.  These ${\cal N}_0 K$ samples are stored and they are 
used to generate additional coarser interactions among the higher-level slices,
\begin{equation} \label{l1b}
(m,m+2)_1=  \langle {\rm sgn} [ (m,m+1)_0 (m+1,m+2)_0 ]
 \rangle_{{\cal P}_0[m+1]}  
= K^{-1} \sum_{m+1} {\rm sgn} [ (m,m+1)_0  (m+1,m+2)_0] \;,
\end{equation}
where the summation $\sum_{m+1}$ extends over the $K$ samples
available for slice $m$.  As will be discussed later on,
the important MLB parameter  $K$ --- subsequently called
{\em sample number} ---  should be chosen as large as possible
to ensure that the second equality in Eq.~(\ref{l1b})
is justified.  The {\em level-1 bonds} (\ref{l1b}) contain precious and
crucial information about the
sign cancellations that occured on the previous level $\ell=0$.
Using these bonds, the density matrix (\ref{rho}) is rewritten as
\begin{equation}
 \rho(P,P)=\sum_{1,2,\ldots,P-1} |(1,2)_0 (2,3)_0 \cdots (P,1)_0| 
 (2,4)_1 \cdots (P-2,P)_1 (P,2)_1 \;.
\end{equation}
Comparing this to Eq.~(\ref{rho}), 
the entire sign problem has been transferred to the next coarser level 
by using the level-1 bonds. 

In the next step, the sampling is carried
out on level $\ell=1$ in order to generate the next-level bonds, 
 i.e.~only slices $m= 2, 6,\ldots, P-2$ are updated,
using the weight ${\cal P}_0 {\cal P}_1$ with 
\begin{equation}
{\cal P}_1 = |(2,4)_1 (4,6)_1 \cdots (P,2)_1| \;.
\end{equation}
Moving the level-1 configurations modifies the level-0 bonds, 
which in turn requires that the level-1 bonds be updated.
A direct re-calculation of these bonds according to Eq.~(\ref{l1b}) 
would be too costly. Instead, the stored configurations 
on level $\ell=0$ are used
to perform an importance sampling of the new level-1 bonds.
Under the test move $m \to m^\prime$, i.e.~$\bbox{r}_m\to \bbox{r}_m^\prime$,
on level $\ell=1$, the bond (\ref{l1b}) can be obtained from 
\begin{equation} \label{eq8}
(m^\prime,m+2)_1 = \frac{\sum_{m+1} \frac{(m^\prime,m+1)_0 (m+1,m+2)_0}
{|(m,m+1)_0 (m+1,m+2)_0|} } {\sum_{m+1} \frac{|(m^\prime,m+1)_0 
(m+1,m+2)_0|} {|(m,m+1)_0 (m+1,m+2)_0|} } \;,
\end{equation}
where $\sum_{m+1}$ runs over the previously
stored MC configurations $\bbox{r}_{m+1}$.
Note that for small values of $K$,
Eq.~(\ref{eq8}) is only approximative, and thus a sufficiently large
value of $K$ should be chosen.
With the aid of Eq.~(\ref{eq8}),  the 
updated level-1 bonds follow with only moderate
computational effort. Generating a sequence of $K$
samples for each slice on level $\ell=1$ and  storing them,
{\em level-2 bonds} can be calculated
in analogy to Eq.~(\ref{l1b}),
\begin{equation}\label{l2b}
(m,m+4)_2 =  \langle {\rm sgn}\left[ (m,m+2)_1
(m+2,m+4)_1 \right] \rangle_{{\cal P}_1 {\cal P}_0} \;.
\end{equation}
Finally, the process is iterated up to the top level $\ell=L$
using the obvious recursive generalization of Eqs.~(\ref{l1b})
and (\ref{l2b}) to define {\em level-$\ell$ bonds}.

Thereby the diagonal elements of the density matrix are obtained as
\begin{equation}\label{rfinal}
\rho(P,P) =\sum_{1,2,\ldots,P-1}
 |(1,2)_0 (2,3)_0 \cdots (P,1)_0| \
 |(2,4)_1 \cdots (P-2,P)_1 (P,2)_1| \cdots
 |(P/2,P)_{L-1} (P,P/2)_{L-1}| \, (P,P)_L \;.
\end{equation}
By virtue of this algorithm, the sign problem is transferred step by step up
to the coarsest level. The expectation value
(\ref{fir}) can thus be computed from 
\begin{equation}\label{expec}
\langle A \rangle = \frac{\langle A(P) \, {\rm sgn}(P,P)_L
\rangle_{\cal P}}{\langle{\rm sgn}(P,P)_L \rangle_{\cal P}} \;.
\end{equation}
The manifestly positive definite MC weight ${\cal P}$
used for the averaging in Eq.~(\ref{expec}) can 
be read off from Eq.~(\ref{rfinal}),
\begin{equation}\label{mcwe}
{\cal P}=|(1,2)_0 (2,3)_0 \cdots (P,1)_0| \
 |(2,4)_1 \cdots (P-2,P)_1 (P,2)_1|  \cdots
 |(P/2,P)_{L-1} (P,P/2)_{L-1}| \, |(P,P)_L|  \;.
\end{equation}
The denominator in Eq.~(\ref{expec}) gives the average sign
and indicates to what extent the sign problem has been solved. 
For proper choice of MLB parameters, in particular the
sample number $K$, this method can solve the sign problem.
The price to pay for the stability of the
algorithm is the increased memory requirement $\sim K^2$ 
associated with having to store the sampled configurations.

\subsection{Real-time simulations} \label{realtime}

The same method, described so far for fermions, can be applied
with minor modifications to a computation of real-time  correlation
functions or occupation probabilities.
For concreteness, let us focus on an equilibrium time-correlation function, 
\begin{equation}\label{corr}
\langle A(0) B(t) \rangle = 
\frac{{\rm Tr} \left\{e^{-(\beta\hbar+it)H/\hbar} A e^{+itH/\hbar} B \right\}}
     {{\rm Tr} \left\{
e^{-(\beta\hbar+it)H/\hbar}   e^{+itH/\hbar} \right\}  }\;.
\end{equation}
Similarly other dynamical properties like the 
thermally symmetrized correlation function \cite{84thi5029},
\begin{equation}\label{scl}
C_s(t) = Z^{-1} {\rm Tr}\{e^{-(\beta\hbar/2+it)H/\hbar} 
A e^{-(\beta/\hbar/2-it)H/\hbar} B\}\;, 
\end{equation}
 with $Z$ being the partition 
function, can be computed.
In terms of path integrals, the traces in (\ref{corr}) involve 
two quantum paths, one propagated backward in time for the duration
$-t$ and the other propagated in complex time for the duration 
$t-i\beta\hbar$. 
Discretizing each of the two paths into $P$ slices, the entire
cyclic path has a total of $2P$ slices. A slice on the first half of them
has length $-t/P$ and on the second half $(t-i\beta\hbar)/P$.
Denoting the quantum numbers (e.g.~spin or position variables)
at slice $j$ by $\bbox{r}_j$, the correlation function (\ref{corr}) reads
\begin{equation}\label{naive}
\frac{ \int d\bbox{r}_1 \cdots d\bbox{r}_{2P}
       B(\bbox{r}_{2P}) A(\bbox{r}_{P}) 
       \prod_{j=1}^{2P} (\bbox{r}_j,\bbox{r}_{j+1})_0 }
     { \int d\bbox{r}_1 \cdots d\bbox{r}_{2P}
       \prod_{j=1}^{2P} (\bbox{r}_j,\bbox{r}_{j+1})_0 }\;,
\end{equation}
where the level-0 bond
$(\bbox{r}_j,\bbox{r}_{j+1})_0$ is again the
short-time propagator between slices $j$ and $j+1$, 
and $\bbox{r}_{2P+1}= \bbox{r}_1$.  
First assign all slices along the discretized path
to different levels $\ell=0,\ldots,L$, where $P=2^L$,
 in close analogy to
the treatment for fermions, see Fig.~\ref{fig0}. 
Each slice $j=1,\ldots,2P$ belongs to a unique level $\ell$, 
such that $j = (2k+1)2^\ell$
and $k$ is a nonnegative integer.  
For instance, slices $j = 1, 3, 5, \ldots$ belong to level $\ell = 0$, 
slices $j = 2, 6, 10, \ldots$ to $\ell = 1$, etc.
The MLB algorithm starts by sampling only configurations
which are allowed to vary on slices associated with the
finest level $\ell = 0$,  using the weight
${\cal P}_0=|(\bbox{r}_1,\bbox{r}_2)_0 \cdots (\bbox{r}_{2P},\bbox{r}_1)_0|$.
The short-time level-$0$ bonds are then employed to
 synthesize longer-time level-$1$
bonds that connect the even-$j$ slices.
Subsequently the level-1 bonds are used
to synthesize level-$2$ bonds, and so on.
In this way the MLB algorithm moves recursively from 
the finest level ($\ell = 0$) up to increasingly coarser levels 
until $\ell = L$, where the measurement
is done using $\bbox{r}_{2P}$ and $\bbox{r}_{P}$.

Generating a MC trajectory containing $K$ samples for each slice 
on level $\ell=0$ and storing these samples, the 
level-1 bonds (\ref{l1b}) can be computed, where the ``sgn'' has
to be replaced by the complex-valued phase $\Phi[z] = e^{i{\rm arg}(z)}$.
Their benefit becomes clear when rewriting
the integrand of the denominator in (\ref{naive}) as
\[
{\cal P}_0 \, \times \, (\bbox{r}_2,\bbox{r}_4)_1 \cdots 
(\bbox{r}_{2P-2},\bbox{r}_{2P})_1 (\bbox{r}_{2P},\bbox{r}_2)_1 \;.
\]
Comparing this to (\ref{naive}),  
the entire sign problem has been transferred to the next coarser level.
In the next step, the sampling is carried
out on level $\ell=1$ in order to compute the next-level bonds, 
using the weight ${\cal P}_0 {\cal P}_1$ with 
${\cal P}_1 = |(\bbox{r}_2,\bbox{r}_4)_1  \cdots (\bbox{r}_{2P},\bbox{r}_2)_1|$.
Generating a sequence of $K$ samples for each slice on level
$\ell=1$, and storing these samples, 
level-2 bonds can be calculated,
\[
(\bbox{r}_j,\bbox{r}_{j+4})_2 =  
\langle {\Phi}\left[ (\bbox{r}_j,\bbox{r}_{j+2})_1
(\bbox{r}_{j+2},\bbox{r}_{j+4})_1 \right]
 \rangle_{{\cal P}_0 {\cal P}_1 } \;.
\]
This process is then  iterated up to the top level.
Finally, the correlation function (\ref{corr}) can be computed from 
\begin{equation}\label{expecta}
\frac{\langle B(\bbox{r}_{2P}) A(\bbox{r}_P) 
 {\Phi}[(\bbox{r}_P,\bbox{r}_{2P})_L (\bbox{r}_{2P},\bbox{r}_P)_L]
\rangle_{\cal P}}
     {\langle{\Phi}[(\bbox{r}_P,\bbox{r}_{2P})_L 
(\bbox{r}_{2P},\bbox{r}_P)_L] \rangle_{\cal P}} \;,
\end{equation}
with the positive definite MC weight 
 ${\cal P} = {\cal P}_0 {\cal P}_1 \cdots {\cal P}_L$.
The denominator in Eq.~(\ref{expecta}) gives the {\em average phase} 
and indicates to what extent the sign problem has been solved. 
With a suitable choice of 
MLB parameters, the average phase remains close to unity 
even for long times.  

\subsection{Accuracy and pitfalls}
\label{performance}

Next questions concerning the {\em exactness} and {\em 
performance}  of the MLB 
approach are addressed, in particular the dependence on the 
sample number $K$.
Clearly, $K$ needs to be sufficiently large
to produce a reliable estimate for the level-$\ell$ bonds. 
If these bonds could be calculated exactly --- corresponding
to the limit $K\to \infty$ ---, the manipulations leading to 
Eq.~(\ref{rfinal}) yield the exact result. Hence for large
enough $K$, the MLB algorithm must become exact 
and completely solve the sign problem. 
Since the level-$\ell$ bonds can however only be computed for finite $K$, 
the weight function ${\cal P}$ amounts to using a  noisy estimator,
which in turn can introduce {\em bias} into the algorithm \cite{kuti}.  
In principle, this problem could be avoided by using a linear
acceptance criterion instead of the algorithmically simpler
 Metropolis choice \cite{qmcgeneral} which was employed in the
applications reported here. 
But even with the Metropolis choice, the bias can be
made arbitrarily small by increasing $K$.
Therefore, with sufficient computer memory, 
the MLB approach can be made to give numerically exact results.  
One might then worry about the actual value of $K>K^*$ 
required to obtain stable and exact results.  If the critical
value  $K^*$ were to scale exponentially with $\beta$ and/or system
size, the sign problem would be present in disguise again.

Although a rigorous non-exponential bound on $K^*$ has not yet
been established, our
experience with the MLB algorithm indicates that
this scaling is at worst algebraic.  This is corroborated by
a recent careful study of this issue \cite{mischa}.
Therefore the {\em exponential severity of the sign problem
is replaced by an algebraic one under MLB}. A heuristic argument
supporting this statement goes as follows.
If one needs $K$ samples for each slice on a given level in order
to have satisfactory statistics despite of the sign problem,
the total number of paths needed in the naive approach
depends exponentially on $P$, namely  $\sim K^P$. 
This is precisely the well-known exponential
severity of the sign problem under the 
naive approach.  However, with MLB the work on the last
level, which is the only one affected by a sign problem provided $K$
was chosen sufficiently large, scales only $\sim K^L$.
Note that it does not scale $\sim K$ because one must update
the level-$\ell$ bonds on all $L$ finer levels as well.
 So in MLB, the work needed to sample the $K^P$
paths with satisfactory statistical accuracy
grows $\sim K^{\log_2 P} = P^{\log_2 K}$, i.e.~algebraically with $P$.
An important point to mention at this point concerns the high-temperature 
(or short-time) limit, where the direct PIMC simulation
does not face a significant sign problem.  
In this case, however, the above-mentioned bias problematics
of the MLB-PIMC is quite serious and can give erroneous results.
Fortunately, since that regime is of little interest to MLB,
this is not a serious restriction.  A more detailed discussion
of this point can be found in Ref.~\cite{mischa}.

To elucidate how the MLB algorithm works in practice, in
Table \ref{tabledot} simulation results for $N=8$ electrons
in a quantum dot 
at various values of $K$ are listed. 
For details, see Sec.~\ref{appl:dot}.
Compared to the naive approach where $K=1$, 
using a moderate $K=200$ already increases the average sign 
from $0.02$ to $0.63$, making it possible to
 obtain more accurate results from much fewer
samples.  The data in Table \ref{tabledot} also confirms that the bias 
can be systematically eliminated by increasing $K$, so that 
the energy found at $K\geq 200$ essentially represents the exact result 
(within error bars).  The value $K^*\approx 200$ is quite
typical for many applications.  For a simple model system,
a value $K^*\approx 50$ was found in Ref.~\cite{mischa}.
Table \ref{tabledot} also shows that
 the CPU time per sample scales linearly with $K$,  
whereas memory requirements grow $\sim K^2$. 

\begin{table}
\caption{\label{tabledot}
MLB results for $N=8$ and $\lambda=2$ (see Sec.~\ref{appl:dot}).
 $N_s$ is the number of samples (in $10^4$),
$t_{\rm CPU}$ the total CPU time (in hours),
 MB the required memory (in mega-bytes), and
$\langle {\rm sgn} \rangle$ the average sign.
Bracketed numbers are error estimates.}  
\begin{tabular}{llllll}\hline
$K$ & $N_s$ &  $t_{\rm CPU}$ & MB  &
$\langle {\rm sgn} \rangle$ & $E_N/\hbar\omega_0$ \\ \hline
1   & 120  &   95 &   2 & 0.02 & $48.6(3)$\\
100 &   7  &   33 &  14 & 0.48 & $48.43(8)$\\
200 &   9  &   83 &  30 & 0.63 & $48.55(7)$\\ 
400 &   8  &  174 &  64 & 0.73 & $48.53(9)$\\
600 &  10  &  308 &  96 & 0.77 & $48.54(8)$\\
800 &   9  &  429 & 150 & 0.81 & $48.59(8)$\\ 
\end{tabular}
\end{table}

\subsection{Effective actions}
\label{effact}

So far the MLB algorithm was only discussed for the case of
nearest-neighbor interactions along the Trotter/time direction.
This situation is encountered under a direct Trotter-Suzuki breakup of the
short-time propagator.
In many cases, however, one has to deal with
effective actions that may include {\em long-ranged
interactions} along the (complex) time direction.  
Such effective actions arise from degrees of freedoms having been traced out, 
e.g.~a harmonic heat bath \cite{weiss},
or through a Hubbard-Stratonovich transformation in 
auxiliary-field MC simulations of lattice fermions \cite{qmcgeneral}.
Remarkably, because such effective
actions capture much of the physics such as symmetries or
the dissipative influence of traced-out degrees of freedom, 
the corresponding path integral very often exhibits a
significantly reduced intrinsic sign problem compared to the original
(time-local) formulation. 
To be specific, let us focus on the dynamical sign problem
arising in real-time PIMC computations here.  
The modifications required to implement the method for
fermion simulations are then straightforward. 

Let us consider a discretized path integral along a slightly
different but fully equivalent contour in 
the complex-time plane compared to Sec.~\ref{realtime}, namely
a forward branch from $t=0$ to $t=t^*$, 
where $t^*$ is the maximum time studied in the simulation, 
followed by a branch going back to the origin, 
and then by an imaginary-time branch 
from $t=0$ to $t=-i\hbar \beta$. Here a
``factorized'' initial preparation is studied,
where the relevant degrees of freedom,
$\bbox{r}(t)$, are held fixed for $t<0$ \cite{weiss}.  
That implies that the imaginary-time dynamics must be
frozen at the corresponding value, and one only needs to 
sample on the two real-time branches.  Note that such
a nonequilibrium calculation cannot proceed 
by first doing an imaginary-time QMC simulation 
followed by a
generally troublesome
analytic continuation of the numerical data \cite{qmcgeneral}.
Using time slices of length $t^*/P$,  
forward [$\bbox{r}(t_m)$] and backward [$\bbox{r}'(t_m)$]
path configurations at time $t_m=m t^*/P$ are combined
into the configuration $\bbox{s}_m$, where $m=1,\ldots,P$.
The configuration at $t=0$ is held fixed,
and for $t=t^*$ one must be in a diagonal state, $\bbox{r}(t^*)
=\bbox{r}'(t^*)$.  For an efficient application of the current 
method, it is essential to combine several neighboring
slices $m$ into new blocks.  For instance, think of $m=1,\ldots,5$
as a new ``slice'' $\ell=1$, $m=6,\ldots,10$ as another slice $\ell=2$,
and so on.  Combining $q$ elementary slices
into a block $\bbox{s}_\ell$, instead of the original $P$ slices one has
$L=P/q$ blocks, where $L$ is the number of MLB levels.
Instead of the ``circular'' structure of the time contour
inherent in the trace
operation, it is actually more helpful
to view the problem as a linear chain, where the 
MLB scheme proceeds from left to right.  
In actual applications, there is considerable freedom in 
how these blocks are defined, e.g.~if 
there is hardly any intrinsic sign problem, or if there
are only few variables in $\bbox{r}$, one may choose
larger values of $q$. 
Additional flexibility can be gained
by choosing different $q$ for different blocks.

Say one is interested in sampling the configurations $\bbox{s}_L$ on the top
level $\ell=L$ according to the appropriate
matrix elements of the (reduced) density matrix,
\begin{equation} \label{rhored}
\rho(\bbox{s}_L) = Z^{-1}\sum_{ \bbox{s}_1,\ldots,
\bbox{s}_{L-1}} \exp\{- S[\bbox{s}_1, \ldots,\bbox{s}_L] \} \;,
\end{equation}
where $S$ is the effective action under study and
$Z$ is a normalization constant so that 
$\sum_{\bbox{s}_L} \rho(\bbox{s}_L) = 1$.
Due to the time-non-locality of this action, there will be interactions
among all blocks $\bbox{s}_\ell$.  The
sum in Eq.~(\ref{rhored}) denotes either an integration over continuous
degrees of freedom or a discrete sum. In the case
of interest here, the effective action is complex-valued and
$e^{-S}/|e^{-S}|$
represents an oscillatory phase factor $\Phi$. 
The ``naive approach'' to the
sign problem is to sample configurations using the 
positive definite weight function
${\cal P} \sim |\exp\{-S\}|$,
and to include the oscillatory phase in the accumulation
procedure.  
Below it is assumed that one can decompose the effective action
according to
\begin{equation} \label{decompos}
S[\bbox{s}_1, \ldots,\bbox{s}_L]  = \sum_{\ell=1}^L
W_\ell[\bbox{s}_\ell,\ldots,\bbox{s}_L] \;.
\end{equation}
All dependence on a configuration $\bbox{s}_\ell$ is then
contained in the ``partial actions'' $W_{\lambda}$ with
$\lambda\leq \ell$.   
One could, of course, put all $W_{\ell>1}=0$,
but the approach becomes more powerful if 
a nontrivial decomposition is possible.  
 
Let us now describe the algorithm in some detail,
employing a somewhat different but equivalent notation than before.
This may be helpful to some readers in order 
to better understand the MLB algorithm, see
also Ref.~\cite{mischa} for a formulation of  
Sec.~\ref{mlbform} in this notation.
The MC sampling starts on the finest level $\ell=1$, where
only the configuration $\bbox{s}_{\ell=1}$ containing the elementary slices
$m=1,\ldots,q$ will be updated with all  $\bbox{s}_{\ell>1}$ remaining fixed
at their initial values $\bbox{s}_\ell^0$.
Using the weight function
\[
{\cal P}_0[\bbox{s}_1] =  |\exp\{-W_1[\bbox{s}_1,
\bbox{s}_2^0, \ldots, \bbox{s}_L^0]\}| \;,
\]
 generate $K$ samples $\bbox{s}_1^{(i)}$,
where $i=1,\ldots,K$,  and store them for later use. 
As usual, the sample number $K$ should be chosen large enough.
For $K=1$, the algorithm will simply reproduce the naive approach.
The stored samples are now employed to generate information about
the sign cancellations.  All knowledge about the interference that occured
at this level is encapsulated in the quantity
\begin{eqnarray} \label{l11} 
B_1 &=& \left\langle\frac{ 
  \exp\{- W_1[\bbox{s}_1, \ldots, \bbox{s}_L]\}
                           }{
  |\exp\{- W_1[\bbox{s}_1, \bbox{s}_2^0, \ldots, \bbox{s}_L^0]\}|
                            } \right\rangle_{{\cal P}_0[\bbox{s}_1]}
 = C_0^{-1} \sum_{\bbox{s}_1} 
\exp\{- W_1[\bbox{s}_1, \ldots, \bbox{s}_L]\}\\
\nonumber
 &=& K^{-1} \sum_{i=1}^K \frac{ 
\exp\{- W_1[\bbox{s}_1^{(i)}, \bbox{s}_2, \ldots, \bbox{s}_L]\}
                                     }{
|\exp\{- W_1[\bbox{s}_1^{(i)}, \bbox{s}_2^0, \ldots, \bbox{s}_L^0]\}|
                                      } 
= B_1[\bbox{s}_2,\ldots,\bbox{s}_L] \;,
\end{eqnarray}
which are analogously called level-1 bonds,
with the normalization constant
$C_0=\sum_{\bbox{s}_1} {\cal P}_0[\bbox{s}_1]$.
Combining the second expression in Eq.~(\ref{l11}) with
Eq.~(\ref{rhored}),  the
density matrix reads
\begin{equation}  \label{rhos}
\rho(\bbox{s}_L) = Z^{-1}\sum_{ \bbox{s}_2,\ldots , \bbox{s}_{L-1}}
\exp\left \{-\sum_{\ell>1}W_\ell \right \} 
C_0 B_1 
 = Z^{-1}\sum_{ \bbox{s}_1,\ldots , \bbox{s}_{L-1}} 
{\cal P}_0 B_1 \prod_{\ell>1} e^{-W_\ell} \;. 
\end{equation}
When comparing Eq.~(\ref{rhos}) with Eq.~(\ref{rhored}), 
the sign problem has now been transferred to
levels $\ell>1$, since oscillatory phase factors only 
arise when sampling on these higher levels.
  Note that  $B_1=B_1[\bbox{s}_2,\ldots,\bbox{s}_L]$
introduces couplings among {\em all} levels $\ell>1$, 
in addition to the ones 
already contained in the effective action $S$.

Next proceed to the next level $\ell=2$ and, according to
Eq.~(\ref{rhos}), update
configurations for $m=q+1,\ldots,2q$ using the weight
\begin{equation}
{\cal P}_1[\bbox{s}_2] =
|B_1[\bbox{s}_2, \bbox{s}_3^0, \ldots, \bbox{s}_L^0]
\exp\{-W_2[\bbox{s}_2, \bbox{s}_3^0, \ldots, \bbox{s}_L^0]\}| \;.
\end{equation}
Under the move $\bbox{s}_2\to \bbox{s}_2'$, one
should then resample and update the level-1 bonds, $B_1\to B_1'$.
Exploiting the fact
that the stored $K$ samples $\bbox{s}_1^{(i)}$ are correctly 
distributed for the original configuration $\bbox{s}_2^0$, 
the updated bond can be computed  according to
\begin{equation}\label{bupdate}
B_1^\prime= K^{-1}\sum_{i=1}^K \frac{ 
\exp\{- W_1[\bbox{s}_1^{(i)}, \bbox{s}_2', \ldots, \bbox{s}_L]\}
                                    }{
|\exp\{- W_1[\bbox{s}_1^{(i)}, \bbox{s}_2^0, \ldots, \bbox{s}_L^0]\}|
                                     } \;.
\end{equation}
Again, to obtain
an accurate estimate for $B_1^\prime$, the number $K$ should be
sufficiently large.
In the end, sampling under the weight ${\cal P}_1$
implies that the probability for accepting the move $\bbox{s}_2\to 
\bbox{s}_2^\prime$
under the Metropolis algorithm is
\begin{equation}
p = \left| \frac{\sum_i  \frac{\exp\{-W_1[\bbox{s}_1^{(i)},\bbox{s}'_2,
\bbox{s}_3^0,
\ldots]\}}{|\exp\{-W_1[\bbox{s}_1^{(i)},\bbox{s}_2^0,\ldots]\}|}}
{\sum_i  \frac{\exp\{-W_1[\bbox{s}_1^{(i)},\bbox{s}_2,\bbox{s}_3^0,
\ldots]\}}{|\exp\{-W_1[\bbox{s}_1^{(i)},\bbox{s}_2^0,\ldots]\}|}} \right |
\times \left| \frac{ \exp\{-W_2[\bbox{s}'_2,\bbox{s}_3^0,\ldots]\}}
{\exp\{-W_2[\bbox{s}_2,\bbox{s}_3^0,\ldots]\}} \right | \;.
\end{equation}
Using this method, one generates $K$ samples 
$\bbox{s}_2^{(i)}$, stores them,
and computes the level-2 bonds, 
\begin{eqnarray} \label{bond2}
B_2 &=& \left\langle \frac{
B_1[\bbox{s}_2, \bbox{s}_3,\ldots]
\exp\{- W_2[\bbox{s}_2, \bbox{s}_3,\ldots]\}
                          }{
|B_1[\bbox{s}_2, \bbox{s}_3^0, \ldots]
\exp\{- W_2[\bbox{s}_2, \bbox{s}_3^0, \ldots]\} |
                           } \right\rangle_{{\cal P}_1[\bbox{s}_2]}
= C_1^{-1} \sum_{\bbox{s}_2} B_1[\bbox{s}_2,\ldots]
 \exp\{- W_2[\bbox{s}_2, \ldots]\} \\ \nonumber
&=& K^{-1} \sum_{i=1}^K \frac{ 
B_1[\bbox{s}_2^{(i)}, \bbox{s}_3, \ldots]
\exp\{- W_2[\bbox{s}_2^{(i)}, \bbox{s}_3, \ldots]\}
                                    }{
|B_1[\bbox{s}_2^{(i)},\bbox{s}_3^0, \ldots]
\exp\{- W_2[\bbox{s}_2^{(i)}, \bbox{s}_3^0, \ldots]\}|
                                     }
= B_2[\bbox{s}_3, \ldots, \bbox{s}_L] \;,
\end{eqnarray}
with $C_1 = \sum_{\bbox{s}_2} {\cal P}_1[\bbox{s}_2]$. 
Following above strategy, one then rewrites the
reduced density matrix by combining Eq.~(\ref{rhos}) and  the 
second expression in Eq.~(\ref{bond2}),
\begin{equation}\label{rhos2}
\rho(\bbox{s}_L) = Z^{-1} \sum_{\bbox{s}_3, \ldots, \bbox{s}_{L-1}}
\exp\left\{ - \sum_{\ell>2} W_\ell \right\}
C_0 C_1 B_2 = Z^{-1} \sum_{\bbox{s}_1, \ldots, \bbox{s}_{L-1}} 
{\cal P}_0 {\cal P}_1 B_2 \prod_{\ell>2} e^{-W_\ell} \;.
\end{equation}
Clearly, the sign problem has been transferred one block further
to the right along the chain. Note that the normalization
constants $C_0, C_1,\ldots$ depend only on the initial configuration
$\bbox{s}_\ell^0$ so that their precise values need not be known.
This procedure is then iterated in a recursive manner.
Sampling on level $\ell$ using the 
weight function
\begin{equation}
{\cal P}_{\ell-1}[\bbox{s}_\ell] =
|B_{\ell-1}[\bbox{s}_\ell, \bbox{s}_{\ell+1}^0, \ldots]
 \exp\{-W_\ell[\bbox{s}_\ell, \bbox{s}_{\ell+1}^0, \ldots]\}|
\end{equation}
requires the recursive update of all bonds $B_{\lambda}$
with $\lambda<\ell$. Starting with $B_1\to B_1'$ and putting $B_0=1$,
this recursive update is done according to
\begin{equation} \label{brec}
B'_{\lambda} = K^{-1}\sum_{i=1}^K \frac{ 
B'_{\lambda-1}[\bbox{s}_\lambda^{(i)}, \bbox{s}_{\lambda+1}, \ldots] \exp\{-
W'_\lambda[\bbox{s}_\lambda^{(i)}, \bbox{s}_{\lambda+1}, \ldots]\}
                                    }{
|B_{\lambda-1}[\bbox{s}_\lambda^{(i)},
\bbox{s}_{\lambda+1}^0, \ldots]\exp\{- 
W_\lambda[\bbox{s}_\lambda^{(i)}, \bbox{s}_{\lambda+1}^0, \ldots]\}|
                                     } \;, 
\end{equation}
where the primed bonds or partial actions depend on $\bbox{s}'_{\ell}$
and the unprimed ones on $\bbox{s}^0_\ell$.
Iterating this to get the updated bonds $B_{\ell-2}$ for all 
$\bbox{s}_{\ell-1}^{(i)}$,   
the test move $\bbox{s}_\ell\to \bbox{s}'_\ell$
is then accepted or rejected according to the probability
\begin{equation} \label{prob}
p = \left| \frac{B_{\ell-1}[\bbox{s}'_\ell, \bbox{s}_{\ell+1}^0, \ldots]
\exp\{-W_\ell[\bbox{s}'_\ell, \bbox{s}_{\ell+1}^0, \ldots]\}}
{B_{\ell-1}[\bbox{s}_\ell, \bbox{s}_{\ell+1}^0, \ldots]
\exp\{-W_\ell[\bbox{s}_\ell, \bbox{s}_{\ell+1}^0, \ldots]\}} \right | \;.
\end{equation}
On this level, 
one again generates $K$ samples $\bbox{s}_{\ell}^{(i)}$, stores them
and computes the level-$\ell$ bonds according to
\[
B_{\ell}[\bbox{s}_{\ell+1},\ldots] = K^{-1}
\sum_{i=1}^K \frac{ 
B_{\ell-1}[\bbox{s}_\ell^{(i)}, \bbox{s}_{\ell+1}, \ldots]
\exp\{- W_\ell[\bbox{s}_\ell^{(i)}, \bbox{s}_{\ell+1}, \ldots]\}
                      }{
|B_{\ell-1}[\bbox{s}_\ell^{(i)}, \bbox{s}_{\ell+1}^0, \ldots]
\exp\{- W_\ell[\bbox{s}_\ell^{(i)}, \bbox{s}_{\ell+1}^0, \ldots]\}|
                       }\;.
\]
This process is iterated up to the top level,
where the observables of interest may be computed.
Since the sampling of $B_{\ell}$ requires the resampling of 
all lower-level bonds, the memory and CPU requirements of the
algorithm laid out here are quite large.  
For $\lambda<\ell-1$, one needs to update $B_{\lambda}\to B'_{\lambda}$ for
all $\bbox{s}_{\ell'}^{(i)}$ with $\lambda< \ell' < \ell$, 
which implies a tremendous amount of computer memory and CPU time,
scaling approximately $\sim K^L$ at the top level. 
Fortunately, an enormous simplification can often be achieved 
by exploiting the fact that the interactions
among distant slices are usually weaker than between near-by
slices.  For instance, when updating level $\ell=3$,
the correlations with the configurations $\bbox{s}_1^{(i)}$
may be very weak, and instead of summing over all $K$ samples
$\bbox{s}_1^{(i)}$ in the update of the bonds $B_{\lambda<\ell}$,
one may select only a small subset.  
When invoking this argument, one should be careful to 
also check that the additional interactions coming from
the level-$\lambda$ bonds with $\lambda<\ell$ are sufficiently
short-ranged.  From the definition of these bonds, this
is to be expected though.

\section{Applications}
\label{appl}

In this section, several different applications of the MLB approach will
be presented.  The first will focus on the equilibrium behavior
of interacting electrons in a parabolic quantum dot, a situation
characterized by a fermionic sign problem.  The two other subsections
then deal with the dynamical sign problem.

\subsection{Quantum dots}
\label{appl:dot}

Quantum dots are solid-state artificial atoms with tunable properties.
Confining a small number of electrons $N$ in a 2D
electron gas in semiconductor heterostructures,
novel effects due to the interplay
between confinement and the Coulomb interaction  have been 
observed experimentally \cite{ashoori,ash3,tarucha}.
For small $N$, comparison of experiments to the generalized Kohn
theorem indicates that the confinement potential is parabolic
and hence quite shallow compared to conventional atoms.
Employing the standard electron gas parameter $r_s$ to
quantify the correlation strength,
only for small $r_s$, a Fermi-liquid picture is applicable.
In the low-density (strong-interaction)
limit of large $r_s$, classical considerations
suggest a Wigner crystal-like phase
with electrons spatially arranged in shells.  We call this a
{\em Wigner molecule} due to its finite extent.  
Of particular interest is the crossover regime between
these two limits, where both single-particle and classical descriptions 
break down, and basically no other sufficiently accurate
method besides QMC is available. 
Exact diagonalization is limited to very small $N$
since one otherwise introduces a
huge error due to the truncation of the Hilbert space.
Hartree-Fock (and related) calculations become unreliable
for large $r_s$ and incorrectly favor spin-polarized states.
Furthermore, density functional calculations can introduce
uncontrolled approximations.  
Regarding QMC, to avoid the sign problem,
the fixed-node approximation 
has been employed by Bolton \cite{bolton}
and later by others \cite{umrigar}.
For $N>5$, typical fixed-node
errors in the total energy are found to be of the order of 10$\%$ \cite{mlb}. 
It is then clear that one should resort to exact methods,
especially when looking at spin-dependent quantities, where often
extremely small spin splittings are found.
After our original studies \cite{mlb,egger99}, other studies using 
the naive PIMC approach were published \cite{borrmann,filinov}.  
These studies are however
concerned with the deep Wigner regime $r_s\gapx 20$ \cite{filinov}, which is
essentially a classical regime without sign problem not further discussed
here,  or employ a special virial estimator \cite{borrmann}
that unfortunately appears to be incorrect
except for fully spin-polarized states \cite{reusch}.
A clean 2D parabolic quantum dot in zero magnetic field is described by
\begin{equation}
H= \sum_{j=1}^N \left(\frac{\vec{p}_j{}^2}{2m^*} +
\frac{m^*\omega_0^2}{2} \vec{x}_j{}^2 \right) +
\sum_{i<j=1}^N \frac{e^2}{\kappa|\vec{x}_i -\vec{x}_j|} \;.
\end{equation}
The electron positions (momenta) are
given by $\vec{x}_j \; (\vec{p}_j)$, their effective mass is $m^*$,
and the dielectric constant is $\kappa$. 
The MLB calculations are carried out at fixed $N$
and fixed $z$-component of the total spin, 
$S=(N_\uparrow- N_\downarrow)/2$.
As a check, the exact solution for $N=2$ has been reproduced.

Here results for the energy,
$E = \langle H \rangle$ (since $H$ is a non-diagonal
operator, two Trotter slices are kept at the top level),
and the {\em spin sensitivity} $\xi_N(r_s)$ will be discussed.  
The latter quantity is useful
to study the crossover from weak to strong correlations, 
\begin{equation} \label{xi}
\xi_N(r_s) \propto
 \sum_{S,S'} \int_0^\infty dy \; y \, | g^{}_S(y)-g^{}_{S'}(y) | \;,
\end{equation}
where the prefactor is chosen to give $\xi_N=1$ for $r_s=0$.
This definition makes use of the spin-dependent 
two-particle correlation function
\begin{equation} \label{cr}
g_S^{}(\vec{x}) = \frac{2\pi l_0^2}{N(N-1)}
\left \langle \sum_{i\neq j=1}^N \delta(\vec{x}-\vec{x}_i+
\vec{x}_j )\right \rangle \;,
\end{equation}
which is isotropic. With 
$y=r/l_0$ prefactors are chosen such that $\int_0^\infty dy \,
y g^{}_S(y)=1$.  The confinement length scale $l_0=\sqrt{\hbar/m^*\omega_0}$
allows the interaction strength to be parametrized by the dimensionless
parameter $\lambda=l_0/a
=e^2/\kappa\omega_0 l_0$, where $a$
is the effective Bohr radius of the artificial atom.
For any given $N$ and $\lambda$,
the density parameter $r_s=r^*/a$ with nearest-neighbor distance
$r^*$ follows by identifying $r^*$ with
 the first maximum in $\sum_S g^{}_S(r)$.
The correlation function (\ref{cr})
is a very sensitive measure of Fermi statistics,
in particular revealing the spin-dependent correlation hole.  
Because interactions tend to wash out the Fermi surface,
the spin sensitivity (\ref{xi}) is 
largest for a Fermi gas, $r_s=0$.
Since for $r_s\to \infty$, one approaches the totally classical limit, 
where $g_S(r)$ is completely spin-independent,
$\xi_N(r_s)$ decays from unity at $r_s=0$ down to
zero as $r_s\to \infty$.  The functional dependence of this decay
provides insight about the crossover phenomenon under study.

\begin{figure}
\epsfysize=12cm 
\epsffile{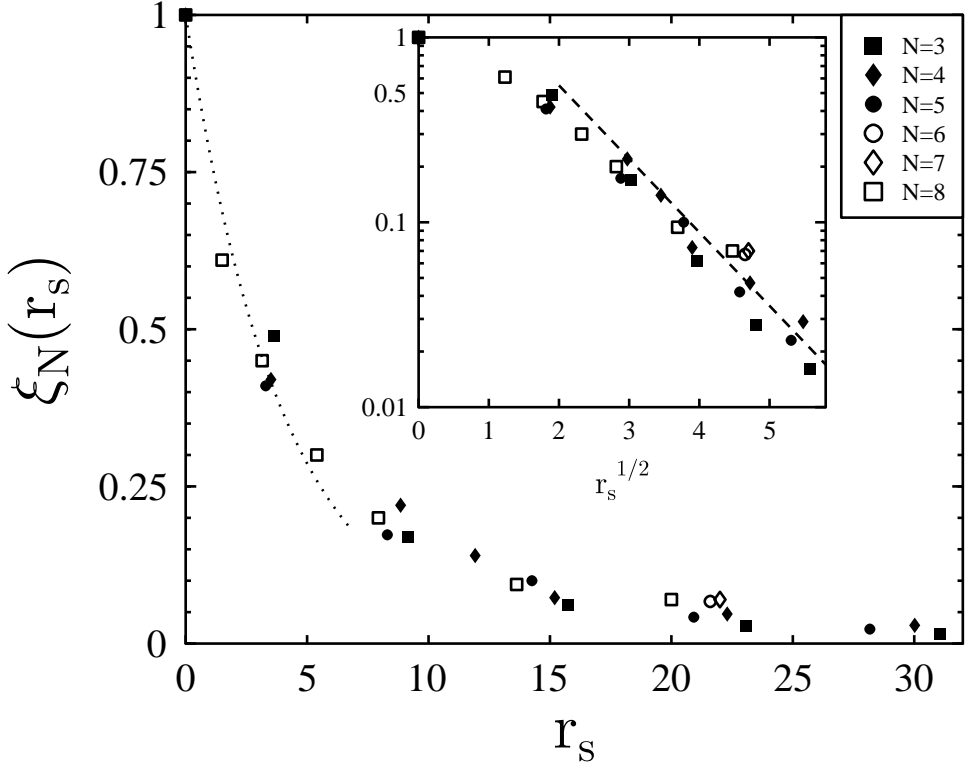}
\caption[]{\label{figxi} Numerical results for $\xi_N(r_s)$
at $k_B T/\hbar \omega_0=0.1$.
Statistical errors are of the order of the symbol size. The
dotted curve, given by $\exp(-r_s/r_c)$ with $r_c=4$, is a guide
to the eye only. The inset shows the same data on a
semi-logarithmic scale as a function of $\sqrt{r_s}$. The
dashed line is the WKB estimate (see text).}
\end{figure}

By computing charge densities, 
the PIMC simulations can directly reveal
 {\sl shell formation} in real space \cite{egger99}. Such a
spatial structure is clear evidence for near-classical
Wigner molecule behavior.
The classical shell filling sequence is as follows.
For $N<6$, the electrons arrange on a ring, but the sixth electron
then goes into the center. 
Furthermore, electrons 7 and 8 enter the outer ring again.
These predictions are in accordance with our PIMC data. 
Clear indications of a spatial shell structure at $N\geq 6$
can be observed already for
$\lambda \approx 4$, albeit quantum fluctuations tend to wash them
out somewhat.  For $\lambda \gapx 4$, charge densities are 
basically insensitive to $S$.  
This is characteristic for a classical Wigner crystal, where
the Pauli principle and spin-dependent properties are
of little importance.  Numerical results for the spin density
in this regime simply follow the corresponding charge density
according to $s_z(r)\simeq (S/N) \rho(r)$. 
A significant $S$-dependence of charge and
spin densities is observed only for weak correlations.
Figure \ref{figxi} reveals that 
$\xi_N(r_s)$ is remarkably
{\sl universal}, i.e.~it depends only very weakly on $N$. Its decay
defines a crossover scale $r_c$, where an exponential
fit for small $r_s$ yields $r_c\approx 4$.
For $r_s>4$, the data can be well fitted by
$\xi(r_s) \sim \exp(-\sqrt{r_s/r_c^\prime})$,
where $r_c^\prime\approx 1.2$.   Remarkably, this is
precisely the behavior expected from a semiclassical WKB estimate
for a Wigner molecule \cite{egger99}.
The crossover value $r_c\approx 4$ is also consistent with the onset of
spatial shell structures in the density, and with 
the spin-dependent ground state energies expected for a Wigner molecule.
Therefore the crossover from weak to strong correlations
is characterized by the surprisingly small value $r_c\approx 4$, instead of
$r_c\approx 37$ found for the bulk 2D electron gas \cite{tanatar}.
This enormous stabilization of the Wigner molecule can be 
ascribed to the effects of the confinement potential. In the 
thermodynamic limit,  $\omega_0\to 0$ with $r_s$ fixed, 
plasmons govern the low-energy physics, and hence
the bulk value $r_c\approx 37$ becomes relevant for very large $N$. 
For GaAs based quantum dots, one can estimate \cite{egger99}
 that for $N\lapx 10^4$,
the value $r_c\approx 4$ is the relevant one.
Remarkably, recent experiments on vertical quantum dots~\cite{ash3}
have found evidence for an even smaller crossover scale $r_c\approx 1.8$.  
The experimental study was carried out in a magnetic
field, and the dot contained several impurities.  Since
both effects tend to stabilize a Wigner crystallized phase,
our prediction and the experimental observation
appear to be consistent.

\begin{table}[t]
\caption{\label{tableen}
MLB data for the energy for various $\{ N, S, \lambda\}$
parameter sets at $k_B T/\hbar \omega_0 = 0.1$. 
Bracketed numbers denote statistical errors.
}
\vspace{0.2cm}
\footnotesize
\begin{tabular}{|llll||llll|} \hline
$N$ & $S$ & $\lambda$ & $E/\hbar\omega_0$&$N$ & $S$ & $\lambda$ &
$E/\hbar\omega_0$\\ \hline
3 & 3/2 & 2 & 8.37(1) & 5 & 5/2 & 8 & 42.86(4) \\
3 & 1/2 & 2 & 8.16(3) &  5 & 3/2 & 8 & 42.82(2) \\
3 & 3/2 & 4 & 11.05(1) & 5 & 1/2 & 8 & 42.77(4) \\
3 & 1/2 & 4 & 11.05(2) & 5 & 5/2 & 10 & 48.79(2) \\
3 & 3/2 & 6 & 13.43(1) & 5 & 3/2 & 10 & 48.78(3) \\
3 & 3/2 & 8 & 15.59(1) & 5 & 1/2 & 10 & 48.76(2) \\
3 & 3/2 & 10 & 17.60(1) & 6 & 3 & 8 & 60.42(2) \\
4 & 2 & 2 & 14.30(5) & 6 & 1 & 8 & 60.37(2) \\
4 & 1 & 2 & 13.78(6) & 7 & 7/2 & 8 & 80.59(4) \\
4 & 2 & 4 & 19.42(1) & 7 & 5/2 & 8 & 80.45(4) \\
4 & 1 & 4 & 19.15(4) &  8 & 4 & 2 & 48.3(2) \\
4 & 2 & 6 & 23.790(12) & 8 & 3 & 2 & 47.4(3) \\
4 & 1 & 6 & 23.62(2) & 8 & 2 & 2 & 46.9(3) \\
4 & 2 & 8 & 27.823(11) & 8 & 1 & 2 & 46.5(2) \\
4 & 1 & 8 & 27.72(1) & 8 & 4 & 4 & 69.2(1) \\
4 & 2 & 10 & 31.538(12) & 8 & 3 & 4 & 68.5(2) \\
4 & 1 & 10 & 31.48(2) & 8 & 2 & 4 & 68.3(2) \\
5 & 5/2 & 2 & 21.29(6) & 8 & 4 & 6 & 86.92(6) \\
5 & 3/2 & 2 & 20.71(8) &  8 & 3 & 6 & 86.82(5) \\
5 & 1/2 & 2 & 20.30(8) & 8 & 2 & 6 & 86.74(4) \\
5 & 5/2 & 4 & 29.22(7) & 8 & 4  & 8  & 103.26(5)  \\
5 & 3/2 & 4 & 29.15(6) & 8 & 3  & 8  & 103.19(4)  \\
5 & 1/2 & 4 & 29.09(6) & 8 & 2 & 8 & 103.08(4)\\
5 & 5/2 & 6 & 36.44(3) &  &  &  &  \\
5 & 3/2 & 6 & 36.35(4) &  &  &  & \\
5 & 1/2 & 6 & 36.26(4) &  &  &  & \\ \hline
\end{tabular}
\end{table}

 MLB results for the energy at different parameter sets $\{N,S,\lambda\}$ are
listed  in Table \ref{tableen}.  For given $N$ and $\lambda$, if the ground
state is (partially) spin-polarized with spin $S$, 
the simulations should yield the same energies for all $S'<S$. 
Within the accuracy of the calculation, this consistency check
is indeed fulfilled.
For strong correlations, $r_s>r_c$, the spin-dependent energy
levels differ substantially from a
 single-particle orbital picture.
In particular, the ground-state spin $S$ can change and
the relative energy of higher-spin states becomes much
smaller than $\hbar\omega_0$.
For $N=3$ electrons, as $r_s$ is increased, a transition occurs from
$S=1/2$ to  $S=3/2$ at an interaction strength
$\lambda\approx 5$ corresponding to $r_s\approx 8$.  
For $N=4$, a Hund's rule case with a small-$r_s$ ground state
characterized by $S=1$ is encountered.
From our data, this standard Hund's rule 
covers the full range of $r_s$.
A similar situation arises for $N=5$, where 
the ground state is characterized by $S=1/2$ for all $r_s$.
Turning to $N=6$, while one has filled orbitals and
hence a zero-spin ground state for weak correlations,
for $\lambda=8$ a $S=1$ ground state is found.  A similar transition
from a $S=1/2$ state for weak correlations
to a partially spin-polarized $S=5/2$ state is 
found for $N=7$. Finally, for $N=8$, 
as expected from Hund's rule,
a $S=1$ ground state is observed for small $r_s$. 
 However, for $\lambda \gapx 4$, corresponding
to $r_s\gapx 10$, the ground state spin changes to $S=2$,
implying a different ``strong-coupling'' Hund's rule.

\begin{figure}
\epsfysize=12cm
\epsffile{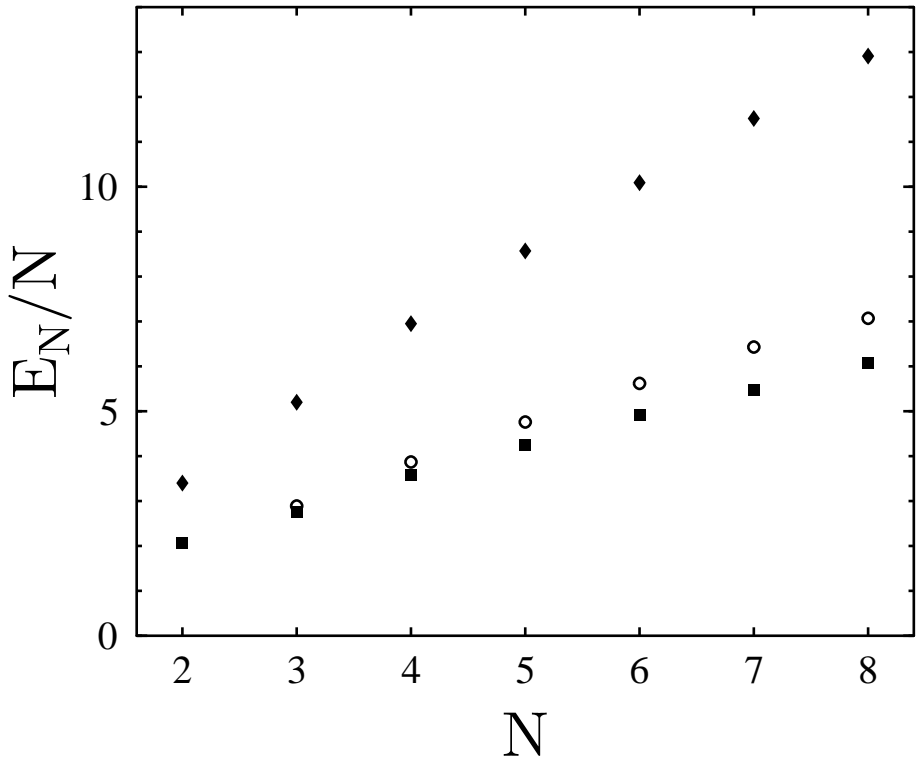}
\caption[]{\label{fig2}
Energy per electron, $E_N/N$, for $S=N/2$ and $k_B T/\hbar\omega_0=1/6$,
in units of $\hbar \omega_0$, for $\lambda=2$ (squares) and 
$\lambda=8$ (diamonds).
Statistical errors are smaller than the symbol size.
Open circles are $T=0$ fixed-node QMC results~\cite{bolton} for 
$\lambda=2$.}
\end{figure}

Let us finally address the issue of {\em magic numbers}. For small $r_s$,
the simple picture of filling up single-particle orbitals predicts
that certain $N$ are exceptionally stable. 
Results for the energy per electron, $E_N/N$, in the spin-polarized state
$S=N/2$, are shown in Figure \ref{fig2}.
Notably, there are no obvious cusps or breaks in the
$N$-dependence of the energy. 
The $\lambda=2$ data in Fig.~\ref{fig2} suggest that an 
explanation of the experimentally
observed magic numbers~\cite{tarucha} has to involve spin and/or 
magnetic field effects, since the
single-particle picture breaks down so quickly.  Remarkably, there are no
pronounced cusps in $E_N/N$ for strong correlations ($\lambda=8$). 
Therefore magic numbers seem to play only a minor role in
the Wigner molecule phase.

\subsection{Real-time simulations: Simple model systems}
\label{appl:simple}

In each of the following examples,  
a time-correlation function was computed directly 
in real time for a simple
model system, with increasing level of complexity.
The average phase is larger than 0.6 for the presented data sets. 

\subsubsection{Harmonic oscillator}

For $H = p^2/2m + m\omega^2 x^2/2$, 
the real and imaginary parts of 
$\langle x(0)x(t) \rangle$ oscillate in time due to vibrational coherence. 
In dimensionless units $m=\omega=1$,
the oscillation period is $2\pi$.
Figure \ref{freal}(a) shows MLB results
for $C(t)={\rm Re}\, \langle x(0) x(t) \rangle$.
With $P=32$ for the maximum time $t=26$, 
$K = 200$ samples were used for sampling the coarser bonds.
Within error bars, the data coincide with the exact result 
and the algorithm produces stable results free of the sign problem. 
Without MLB, the signal-to-noise ratio was practically zero 
for $ t > 2$.

\subsubsection{Two-level system}

For a symmetric two-state system, 
$H = -\frac{1}{2} \Delta \sigma_x$, 
 the dynamics is controlled by tunneling.
The spin correlation function $\langle \sigma_z(0)\sigma_z(t) \rangle$ 
exhibits oscillations indicative of quantum coherence.  
Figure \ref{freal}(b) shows MLB results for
 $C(t)={\rm Re}\,\langle \sigma_z(0)\sigma_z(t) \rangle$,
Putting $\Delta=1$, the tunneling oscillations have a period of $2\pi$.
With $P = 64$ for the maximum time $t=64$,
only $K = 100$ samples were used for sampling the coarser bonds.
The data agree well with the exact result.  
Again the simulation is stable and free of the sign problem. 
Without MBL, the simulation failed for $t > 4$.

\subsubsection{Double-well potential}

Next, let us examine a double-well system 
with the quartic potential $V(x) = -x^2 + \frac{1}{4}x^4$.
At low temperatures, interwell transfer occurs through tunneling motions 
on top of intrawell vibrations.  These two effects
combine to produce nontrivial structures in the position correlation
function.  At high temperatures,
interwell transfer can also occur by classical barrier crossings.
Figure \ref{freal}(c) shows MLB results
for $C(t)={\rm Re}\,\langle x(0)x(t)\rangle$.
The slow oscillation corresponds to interwell tunneling,
with a period of approximately 16.  The higher-frequency 
motions are characteristic of intrawell oscillations.  
In this simulation, $K = 300$ samples were used.
The data reproduce the exact result well, capturing
all the fine features of the oscillations.  
Again the calculation is stable and free of the sign problem,
whereas a direct simulation failed for $t>3$.

\subsubsection{Multidimensional tunneling system}

\begin{figure}
\epsfysize=16cm
\epsffile{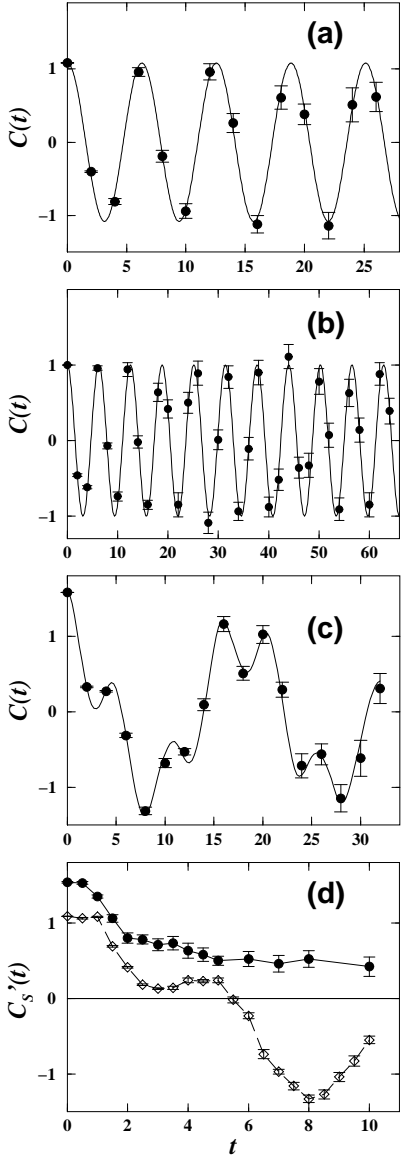}
\caption[]{\label{freal}
MLB results (closed circles) for various systems.
Error bars indicate one standard deviation. 
(a) $C(t)$ for a harmonic oscillator at $\beta\hbar = 1$. 
The exact result is indicated by the solid curve. 
(b) Same as (a) for a two-level system at $\beta\hbar=10$.
(c) Same as (a) for a double-well system at $\beta\hbar=1$.
This temperature corresponds to the classical barrier energy.
(d) $C'_s(t)$ for a double-well system coupled to two
 oscillators at $\beta\hbar = 1$.
For comparison, open diamonds are for the uncoupled $(\alpha=0)$ system.
Note that $C'_s(t)$ is similar but not identical to $C(t)$ shown
in (c). Solid and dashed lines are guides to the eye only.}
\end{figure}

As a final example,  consider a problem with three degrees of freedom, 
in which a particle in a double-well potential is bilinearly coupled to 
two harmonic oscillators.
The quartic potential in the last example is used for the double-well, 
and the harmonic potential in the first example is used for both oscillators.
The coupling constant between each oscillator and the tunneling particle 
is $\alpha = 1/2$ in dimensionless units.
For this example,  the correlation function $C_s(t)$ 
in Eq.~(\ref{scl}) has been computed
for the position operator of the tunneling particle.
Direct application of MC sampling to $C_s(t)$
has generally been found unstable for $t>\beta\hbar/2$
\cite{84thi5029}.
In contrast, employing only moderate values of $K=400$ to 900
 allow to go up to $t=10\beta\hbar$. 
 Figure \ref{freal}(d) shows MLB results
for $C'_s(t)={\rm Re} \,C_s(t)$.
For the coupled system, the position correlations have lost the coherent 
oscillations and instead decay monotonically with time. 
Coupling to the medium clearly damps the coherence and tends to
localize the tunneling particle.

\subsection{Spin-boson dynamics}
\label{appl:sb}

Finally, to demonstrate the performance of the MLB approach for
effective-action-type problems, 
the real-time dynamics of the celebrated spin-boson model \cite{weiss}
\begin{equation} \label{spbos}
H = - (\hbar\Delta/2)\, \sigma_x + (\hbar\epsilon/2)\, \sigma_z  
+ \sum_\alpha
\left[ \frac{p_\alpha^2}{2m_\alpha} +
{\textstyle \frac{1}{2}}  m_\alpha \omega_\alpha^2
\left(x_\alpha - \frac{c_\alpha}{m_\alpha \omega_\alpha^2} \sigma_z\right)^2
\right] 
\end{equation}
is discussed.  This model has a number of important 
applications \cite{weiss}, e.g.~the Kondo problem, 
interstitial tunneling in solids, 
quantum computing and electron transfer
reactions, to mention only a few. 
The bare two-level system (TLS) has a tunneling matrix element $\Delta$
and the asymmetry (bias) $\epsilon$ between the two localized energy levels  
($\sigma_x$ and $\sigma_z$ are 
Pauli matrices). Dissipation is introduced via a linear heat bath, 
i.e.~an arbitrary collection of harmonic oscillators $\{x_\alpha\}$
bilinearly coupled to $\sigma_z$.
Concerning the TLS dynamics, all information about the coupling 
to the bath is contained in the spectral density
$J(\omega)=(\pi/2) \sum_\alpha (c^2_\alpha/m_\alpha\omega_\alpha)\,
\delta (\omega-\omega_\alpha)$,
which has a quasi-continuous form in typical condensed-phase applications
and dictates the form of 
the (twice-integrated) bath correlation function 
\begin{equation}
Q(t) = \int_0^\infty \frac{d\omega}{\pi \hbar} 
\frac{J(\omega)}{\omega^2}
\, \frac{\cosh[\omega\hbar\beta/2]-
\cosh[\omega(\hbar\beta/2-it)]}{\sinh[\omega\hbar\beta/2]}\;.
\end{equation}
For the calculations here, an ohmic spectral density 
$J(\omega) =2\pi\hbar\alpha \omega \exp(-\omega/\omega_c)$
has been assumed,
for which $Q(t)$ can be found in closed form \cite{weiss}.  Here
$\omega_c$ is a cutoff frequency, and the damping strength is 
measured by the dimensionless Kondo parameter $\alpha$. 
In the {\sl scaling limit}\ $\Delta\ll \omega_c$ with
 $\alpha<1$,
all dependence on $\omega_c$ enters via a renormalized
 tunnel splitting \cite{weiss}
\begin{equation}
\Delta_{\rm eff} 
= [\cos(\pi \alpha)\Gamma(1-2\alpha)]^{1/2(1-\alpha)}
(\Delta/\omega_c)^{\alpha/(1-\alpha)}  \Delta \;,
\end{equation}
and powerful analytical and alternative numerical methods 
are readily available \cite{weiss}. 
However, there are important applications (e.g.~electron transfer reactions)
that require to study the spin-boson problem away from 
the scaling limit.  Here one generally 
has to resort to numerical methods.  
Basically all other available computational techniques
can only deal with equilibrium quantities, 
or explicitly introduce approximations; for an overview and 
references, see Ref.~\cite{weiss}.
The MLB approach is computationally more expensive than other methods 
but at the same time unique in yielding numerically
exact results for the nonequilibrium spin-boson dynamics for 
arbitrary system parameters $\Delta, \epsilon, J(\omega)$ and
$\beta=1/k_B T$.

\begin{table}
\caption[]{\label{tablesb}
MLB performance for $\alpha=1/2$, $\omega_c/\Delta=6$, 
$\Delta t^*=10$, $P=40$, and several $L$.
$q_\ell$ denotes the number of slices for $\ell=1,\ldots L$. }
\begin{tabular}{llll}
$K$   & $L$ & $q_\ell$        & $\langle {\rm sgn} \rangle$ \\ \hline
1     & 1   & 40              & 0.03  \\   
200   & 2   & 30 - 10         & 0.14  \\
800   & 2   & 30 - 10         & 0.20  \\
200   & 3   & 22 - 12 - 6     & 0.39  \\
600   & 3   & 22 - 12 - 6     & 0.45  \\
\end{tabular}
\end{table}

Below only results for the {\em occupation probability}
$P(t) =  \langle \sigma_z(t) \rangle$ 
under the nonequilibrium initial preparation $\sigma_z(t<0)=+1$ are
presented.
$P(t)$ gives the time-dependent difference of the 
quantum-mechanical occupation probabilities of the left and right states,
with the particle initially confined to the left state.
To obtain $P(t)$ numerically, in a discretized
path-integral representation one 
traces out the bath to get a long-ranged effective action, the 
{\em influence functional} \cite{weiss}.  In discretized form
the TLS path is represented by spins $\sigma_i,\sigma'_i=\pm 1$
on the forward- and backward-paths, respectively.
The total action $S$ consists of three terms.
First, there is the ``free'' action $S_0$ determined by
 the bare TLS propagator $U_0$, 
\begin{equation}
\exp(-S_0)= \prod_{i=0}^{P-1} U_0(\sigma_{i+1},\sigma_i;t^*/P)
\; U_0 (\sigma'_{i+1},\sigma'_i;-t^*/P) \;,
\end{equation}
where $t^*$ is the maximum time and $P$ the Trotter number.
Next there is the influence functional, $S_I= S_I^{(1)} + S_I^{(2)}$, which 
contains the long-ranged interaction among the spins,
\[
S_I^{(1)} = \sum_{j\geq m} (\sigma_j-\sigma'_j) 
\Bigl\{ L'_{j-m} \, (\sigma_m-\sigma'_m) + i L^{''}_{j-m}\,
(\sigma_m+\sigma'_m) \Bigr\} \;,
\]
where $L_j=L'_j+iL_j^{''}$ is given by 
\begin{equation}
L_{j}  =  [Q((j+1)t^*/P)+Q((j-1) t^*/P) - 2Q(jt^*/P) ]/4
\end{equation}
for $j>0$, and $L_{0}=Q(t^*/P)/4$. 
In the scaling regime at $T=0$, this effective action produces
interactions $\sim \alpha/t^2$ between the spins
(``inverse-square Ising model'').  
The contribution  
\[
S_I^{(2)} = i(t^*/P) \sum_m  \gamma(m t^*/P) (\sigma_m-\sigma'_m) 
\]
describes the interactions with 
the imaginary-time branch where $\sigma_z=+1$,
with the damping kernel 
\[
\gamma(t)= \frac{2}{\pi \hbar} \int_0^\infty
d\omega \frac{J(\omega)}{\omega} \,\cos(\omega t) \;.
\]
The most difficult case for PIMC corresponds to an unbiased
two-state system at zero temperature, $\epsilon=T=0$.
To check the code, the case $\alpha=1/2$ was studied in some detail, where 
the exact solution \cite{weiss} is very simple, 
$P(t) = \exp(-\Delta_{\rm eff} t)$.
This exact solution  only  holds in the scaling limit, which is
already reached for $\omega_c/\Delta=6$ where 
 the MLB-PIMC simulations yield precisely this result.
Typical MLB parameters 
and the respective average sign are listed in Table~\ref{tablesb}.
The first line in Table~\ref{tablesb} corresponds to the naive 
approach. It is then clear that the average sign and
hence the signal-to-noise ratio can
be dramatically improved thus allowing for a study of
long timescales $t^*$.
For a fixed number of levels $L$, the average sign 
grows by increasing the parameter $K$.  Alternatively, for  
fixed $K$, the average sign increases with $L$.  Evidently,
the latter procedure is more efficient in curing the sign problem,
but at the same time computationally expensive.  In practice,
it is then necessary to find a suitable compromise.

\begin{figure}
\epsfysize=10cm
\epsffile{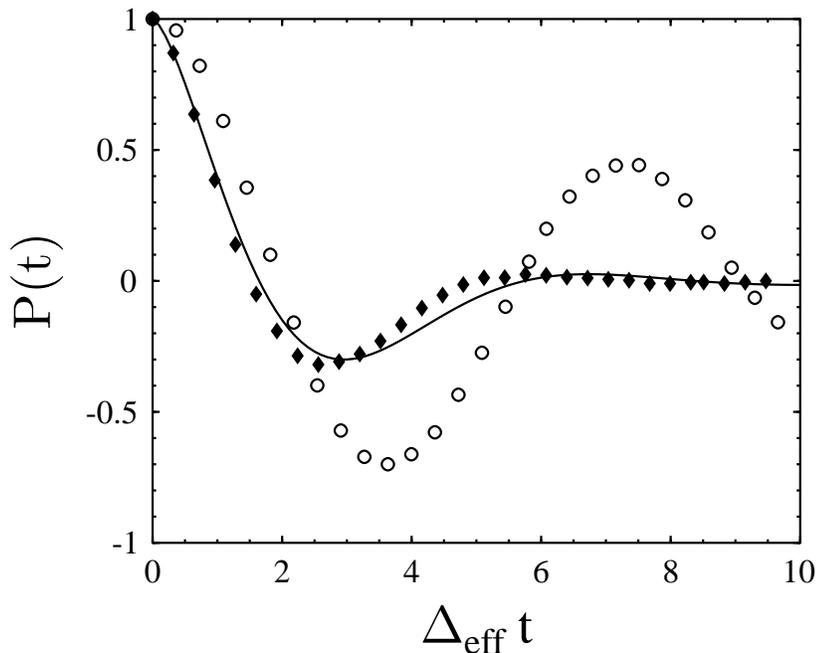}
\caption[]{\label{fsb} Scaling curves for
$P(t)$ for $\alpha=1/4$ with
$\omega_c/\Delta=6$ (closed diamonds) and  
$\omega_c/\Delta=1$ (open circles).  The solid curve is the
NIBA prediction. 
Statistical errors are of the order of the symbol sizes.}
\end{figure}

Figure \ref{fsb} shows scaling curves for $P(t)$ at $\alpha=1/4$ for 
$\omega_c/\Delta=6$ and $\omega_c/\Delta=1$.  The first value
for $\omega_c/\Delta$ is within the scaling regime.
This is confirmed by a comparison to
the noninteracting blip approximation (NIBA)
\cite{weiss}, which is known to be very accurate for $\alpha<1$
in the scaling limit.
However, for $\omega_c/\Delta=1$, scaling concepts
and also NIBA are expected to fail dramatically. 
This is seen in the simulations.
MLB results show that away from the scaling limit,
quantum coherence is able to persist for much longer, and both frequency and 
decay rate of the oscillations differ significantly from the predictions
of NIBA.  In electron transfer reactions in the  
adiabatic-to-nonadiabatic crossover regime, 
such coherence effects can then strongly influence the low-temperature
dynamics.  One obvious and important consequence of 
these coherence effects is the
breakdown of a rate description, implying that theories based on 
an imaginary-time formalism might not be appropriate in this regime.
A detailed analysis of this crossover regime
using MLB is currently in progress \cite{lothar}.

\section{Concluding remarks}
\label{conc}

These notes summarize our previous activities using
the multilevel blocking approach to the sign problem
in path-integral Monte Carlo simulations.  
The approach holds substantial promise towards
relieving (and eventually overcoming) the sign
problem, but clearly there is still much room
for improvement.  The applications presented here
demonstrate unambiguously that there are general
and powerful handles to relieve the sign problem,
even though a problem characterized by an intrinsic
sign problem is still much harder than one without.
We hope that especially young researchers will be
attracted to work on this subject themselves.

\acknowledgments
Parts of this review are based on work with Lothar
M\"uhlbacher.  This research has been supported by the
Volkswagen-Stiftung (I/73 259) and by the
National Science Foundation
(CHE-9970766).

\end{document}